\documentclass[sigconf]{acmart}
\AtBeginDocument{%
  }

\begin{document}

\title{Hyperion: An AI-powered HPC cluster for sciences and humanities research that utilizes ML for predicting job turnaround time}

\author{Jun Zhou}
\email{zhouj@mailbox.sc.edu}
\orcid{0000-0003-0449-8953}
\correspondingauthor
\affiliation{%
  \institution{University of South Carolina}
  \city{Columbia}
  \state{South Carolina}
  \country{USA}
}

\author{Nathan Elgar}
\affiliation{%
  \institution{University of South Carolina}
  \city{Columbia}
  \country{USA}}
\email{elgar@mailbox.sc.edu}

\author{Tawnee' Benedetto}
\affiliation{%
  \institution{University of South Carolina}
  \city{Columbia}
  \country{USA}}
\email{msbthewordguru@gmail.com}

\author{John Richards}
\affiliation{%
  \institution{University of South Carolina}
  \city{Columbia}
  \country{USA}}
\email{richards@mailbox.sc.edu}

\author{Ming Hu}
\affiliation{%
  \institution{University of South Carolina}
  \city{Columbia}
  \country{USA}}
\email{hu@sc.edu}

\author{Greg Wilsbacher}
\affiliation{%
  \institution{University of South Carolina}
  \city{Columbia}
  \country{USA}}
\email{GREGW@mailbox.sc.edu}



\author{Lawrence Miao}
\affiliation{%
  \institution{Boston University}
  \city{Boston}
  \country{USA}
}
\email{miaol2@bu.edu}

\author{Paul Sagona}
\affiliation{%
  \institution{University of South Carolina}
  \city{Columbia}
  \country{USA}}
\email{SAGONA@sc.edu}

\renewcommand{\shortauthors}{Zhou et al.}

\begin{abstract}
Hyperion is an innovative high-performance computing (HPC) cluster developed for researchers in both science and humanities disciplines at the University of South Carolina (USC). Our approach involved constructing a HPC cluster designed to meet the current research needs while accommodating future expansion. Additionally, we developed and trained two machine learning (ML) models to predict turnaround time, including wait time and wall time, and seamlessly integrated them into the Slurm job submission. Finally, we showcase a variety of sample applications hosted on the Hyperion platform.
\end{abstract}

\ccsdesc[500]{Computing methodologies~Object detection}
\ccsdesc[500]{Computing methodologies~Classification and regression trees}
\ccsdesc[500]{Computer systems organization~Heterogeneous (hybrid) systems}
\ccsdesc[500]{Computer systems organization~Special purpose systems}


\maketitle

\section{Introduction}
High-performance computing (HPC) plays a critical role in advancing scientific research by facilitating sophisticated modeling, simulation, and data analysis. HPC systems provide the computational power and analytical capabilities necessary for conducting intricate experiments, processing large datasets, and validating theoretical frameworks that may be impractical to test through traditional experimentation methods. Parallelism and co-processing speed up computations, making it possible to process these large datasets and run massive-scale experiments in less time. Accelerators, such as GPUs, have proven to be crucial in providing advanced processing capabilities, especially in the area of Artificial Intelligence (AI). As a result, AI-related algorithms, such as neural network models, can be more effectively processed. This is significant because AI methodologies have recently been widely applied to various fields of research, including AI itself. 

Most research institutes and scientific labs have built their HPC systems tailored to their own needs, such as bridges 2 GPU clusters at Pittsburgh University\footnote{https://www.psc.edu/resources/bridges-2/}, Stampede 2 at Texas University\footnote{https://tacc.utexas.edu/systems/stampede2/}, Jetstream 2 at Indiana University\footnote{
https://jetstream-cloud.org/index.html}, Anvil cluster at Purdue University\footnote{https://www.rcac.purdue.edu/compute/anvil}, Hive at Georgia Tech\footnote{https://pace.gatech.edu/new-hpc-cluster-hive}, and more. 

\begin{figure}[h]
  \centering
  \includegraphics[width=\linewidth]{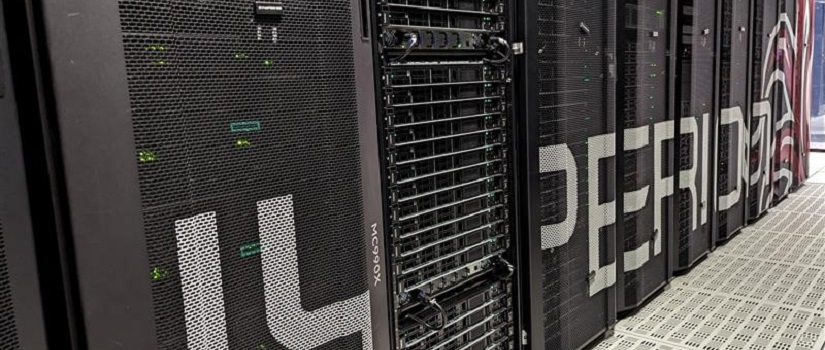}
  \caption{The Hyperion HPC cluster at the University of South Carolina.}
  \label{fig:hyperion}
\end{figure}

The University of South Carolina (USC) joined this trend decades ago. Initially, the model was highly decentralized and efforts were siloed across the university leading to a lack of efficiency in access, scale, and utilization. Then in 2013, several research cyberinfrastructure groups consolidated under the central Information Technology division, combining both staff and technology for a broader support model and a shared HPC facility for the entire university. Centralizing these services increased efficiency by reducing redundancy, establishing standardization of procedures and protocols, realizing cost savings through economies of scale, allowing for enhanced security, and improving access to resources. The success of this centralization then led to broader support from the campus administration and secured funding from various sources including individual researchers (via grants), departments, colleges, and even the Provost. In 2017, USC built the largest HPC cluster to date at the university: Hyperion. As its photo shown in Fig.~\ref{fig:hyperion}, the Hyperion cluster is a heterogeneous system designed to address the computational requirements which were determined through surveying the entire research landscape at USC. The initial configuration consisted of CPU, GPU, large-memory nodes, and a robust high-speed interconnect that could easily grow as needed. Continued investment has expanded this state-of-the-art facility in several phases to accommodate the steady growth in users and has become an efficient core resource for USC researchers. The current Hyperion configuration is outlined in section~\ref{sec:architecture}. The next expansion phase, slated to be online in fall of 2024, consists of CPU expansions and highly desirable GPUs for AI workloads. Hyperion is managed by the Research Computing group under the Division of Information Technology (DoIT) and supports over 759 researchers. It has executed over 15.3 million jobs since its birth with an average wait time of 3.36 hours.

While providing user support to researchers, a question was constantly raised: what is my job turnaround time? To address this question, we trained two new machine learning regression models to predict wait time and wall time (run time) for each newly submitted job. Compared to related previous work conducted by other researchers~\cite{tanash2021ampro}, our models have achieved an average R-squared value of 0.83 for wall time and 0.74 for wait time, demonstrating significantly improved prediction performance. We presented the prediction results to researchers by integrating these two models into Slurm Workload Manager (Slurm)~\footnote{https://slurm.schedmd.com/overview.html} job submission scripts.

Our efforts have been dedicated to supporting researchers in leveraging the capabilities of Hyperion through system setup, personalized consultation, countless workshop series, and grant writing and management assistance. Researchers across various disciplines including engineering, sciences, and humanities, have begun utilizing Hyperion for their endeavors. In the subsequent sections we showcase the significant contributions of three prominent research groups: one each from psychology, material sciences, and
library sciences, illustrating how Hyperion has benefitted their respective research pursuits.

In this paper, we outline the following contributions: 
\begin{itemize}
\item Designed and implemented a hybrid HPC cluster.
\item Developed and trained of two novel models for predicting wall time and wait time.
\item Presented diverse applications hosted on the Hyperion platform.
\end{itemize}

\section{System Architecture}
\label{sec:architecture}

All of the USC Research Computing resources, including HPC clusters, storage, and core network infrastructure, are housed within USC’s DoIT data center, which offers enterprise-level monitoring, cooling, and backup power. The flagship HPC resource, the Hyperion cluster, serves faculty, staff, and students across all domains at USC. Operated on the ``Condo'' model~\cite{bottum2013condo}, users who purchase hardware within the cluster enjoy subsidized data center infrastructure in exchange for donating their idle cycles back to the USC community via free, shared queues. These resources are managed and automated via the Slurm Workload Manager.

Initially configured with 240 total nodes running 28-core Intel Xeon E5-2680 CPUs, these nodes were divided into a CPU-only processing queue of 224 nodes, 8 large-memory nodes with 1.4TB of RAM, and 8 GPU nodes with dual NVIDIA P100 GPUs. This modest complement of GPU nodes provided users with a development environment for GPU workloads that could be scaled up on larger national HPC cluster resources. However, demand for additional GPU processing resources surged over the next two years, surpassing the available GPU resources. To meet the needs of both the growing standard CPU compute user base and the increasing demand for machine learning tasks, the first hardware refresh in 2019 added 120 CPU compute nodes running 48-core Intel Xeon Platinum 8260 CPUs and 44 GPU nodes with dual NVIDIA V100 GPUs.

Hyperion currently comprises 291 standard CPU compute nodes, 44 GPU nodes, and 8 Big Memory nodes, providing a total of 16,616 CPU cores. The Compute and GPU nodes have 128-256 GB of RAM, while the Big Memory nodes have 2TB of RAM. All nodes are equipped with EDR InfiniBand (100 Gb/s) interconnects and have access to 1.4 PB of GPFS storage. The most recent expansion of GPU nodes brought Hyperion's total single precision performance to 1,232 teraFLOPS (TFLOPS).

The latest hardware upgrade was completed in May 2023. This added 86 standard CPU nodes and eight 2TB RAM large memory nodes, running 64 core Intel Xeon Platinum 8358 CPUs. Concurrent with this upgrade, we retired the open first-generation 28 core standard compute nodes and all of the first-generation large memory nodes. This allowed us to recoup the less efficient nodes’ power and network infrastructure footprint for the newer, more efficient compute nodes. All of Hyperion’s nodes are connected via a high-speed, low latency, EDR InfiniBand network at 100 Gbit. This high-speed fabric links Hyperion to a 1.4 petabyte high-performance GPFS scratch filesystem, with additional 100 Gbit Ethernet links to on-premises research storage via Globus Data Transfer Nodes. 

In addition to the larger expansions, specialized systems like an Nvidia DGX A100 have been added to the overall Hyperion environment. In October of 2020, Research Computing partnered with the USC AI Institute to purchase an NVIDIA DGX A100, which was installed in December of that year, adding 8x A100 40gb GPUs that offer a combined 5 petaFLOPS (PFLOPS) of AI performance.

In fall 2023, USC was awarded a National Science Foundation’s Major Research Instrumentation (NSF MRI) grant in order to expand the AI capabilities of USC and neighboring Minority-Serving Institutions (MSI). The instrument will have a focus on GPU scaled workloads and will provide infrastructure for training in AI to underserved institutions across South Carolina. The configuration of this expansion consists of 28 standard CPU compute nodes with dual Intel 56 core CPUs, 9 GPU nodes with 4x A100 GPGPUs, and 1 GPU node with 4x H100 GPGPUs. The additional hardware will be installed and available to researchers in the fall of 2024.

%

\section{Turnaround Time Prediction}
\label{sec:prediction}

Estimating job turn-around time is essential for researchers to effectively plan their work and utilize HPC resources efficiently. While many institutes have developed their own estimation methods, these models are often tailored to their specific resources and may not yield satisfactory results when applied to different systems like Hyperion. For instance, Table~\ref{tab:wait-time}, Row 3 presents the outcomes based on similar methods proposed in~\cite{tanash2021ampro}, which exhibit low prediction accuracy (indicated by low R-squared values).

To address this challenge, we developed two new machine learning models using historical job data from XDMod~\cite{palmer2015open}: one for predicting wait time and another for predicting wall time. From the 15.3 million jobs submitted through Slurm on Hyperion, we constructed a dataset that comprised all jobs recorded in the XDMod data warehouse export from 2018 to 2023, totaling 4.9 million jobs across 704 users. We partitioned the dataset into 70\% for training and 30\% for testing. Notably, unlike previous studies, we did not perform data cleaning on these data, resulting in more realistic models.

We experimented with various machine learning regressors for wait time and wall time prediction and found that the Random Forest~\cite{rigatti2017random} regressor yielded the best performance. We conducted extensive feature selection and tested different combinations of features. The ``Submit Time'', represented as a timestamp and normalized to midnight on January 1st of each year, was clearly the most influential feature in both models, indicating its crucial role in determining job turnaround time.

Table~\ref{tab:wait-time} presented the prediction results on the average outcomes over the year 2018 through 2023, showing in R-squared values. A higher R-squared value indicates a better model performance. We observed that for wait time prediction, the feature set \{"Submit Time, Nodes, Cores, GPUs, Queue ID"\} achieved the highest prediction accuracy. Conversely, for wall time prediction, the feature set \{"Submit Time, Nodes, Cores, GPUs, User ID, Queue ID"\} demonstrated the highest prediction accuracy.

By analyzing the importance of each feature in the trained models, we found that the ``Submit Time'' feature had the highest value, followed by ``User ID''. This suggests that for clusters like Hyperion, which primarily handle short-duration jobs, turn-around time is predominantly influenced by the submission time. For instance, jobs submitted during off-peak hours or on days with fewer paper or grant deadlines tend to execute more promptly.

Given that we register 2 to 3 new users each week, we have opted to utilize the model without ``User ID'' in our prediction, despite the slightly better accuracy on wall time when it is  incorporated. Both models have been integrated into Slurm job submission scripts, enabling users to receive notifications regarding their job turnaround time immediately after submission.

\begin{table}[htbp]
    \centering
    \caption{Average R-squared values for wait time and wall time across the years 2018 to 2023}
    \begin{tabular}{ |p{8.5cm}|p{2cm}|p{2cm}| }
        \hline
        Features&Wait Time &Wall Time \\
        \hline
        Submit Time, Nodes, Cores, GPUs, User ID, Queue ID & 0.8665& \textbf{0.7459}\\
   Submit Time, Nodes, Cores, GPUs, Queue ID  & \textbf{0.8667}& 0.7218\\
   Nodes, Cores, GPUs, User ID, Queue ID  & 0.3270& 0.3457\\
   
        \hline
    \end{tabular}
    
    \label{tab:wait-time}
\end{table}

\section{Applications}
\label{sec:applications}

\subsection{Phonon Property Prediction of Crystals}

The advent of machine learning and artificial intelligence has revolutionized many aspects of modern science and technology and has sparked significant interest in the material science community in recent years. Despite some early deployment of  ML/AI in thermal science area, the power of AI has not been maximized. Existing ML methods for predicting phonon properties of crystals are limited to either small amount of training data or a material-to-material basis, primarily due to the exponential scaling of model parameters with the number of atomic species or elements. This renders high-throughput infeasible when facing large-scale new materials. 

\begin{figure*}[h]
  \centering
  \includegraphics[width=0.9\linewidth]{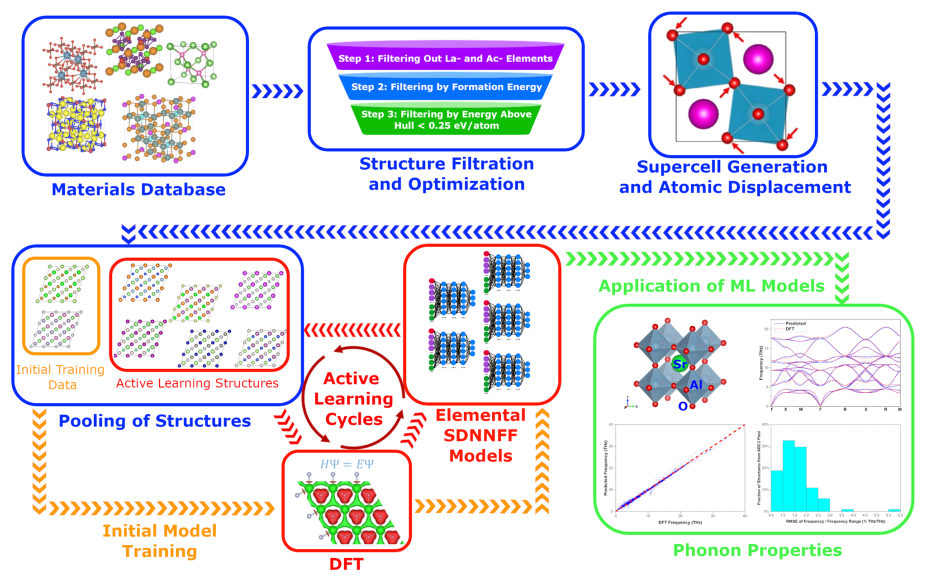}
  \caption{Workﬂow for training Elemental-SDNNFF models.}
  \label{fig:SDNNFF}
\end{figure*}

The research activities in Prof. Ming Hu’s group using Hyperion cluster include high-throughput first principles calculations of physical properties of inorganic crystals and development of the state-of-the-art ML/AI approaches for predicting energy carriers' transport behaviors in those inorganic materials~\cite{rodriguez2023unlocking}~\cite{ rodriguez2023million}. Both traditional ML methods (such as random forest) and novel graph neural networks are deployed for training, testing, and prediction of lattice vibrations (phonons) and relevant properties. Particular focus is our recently developed Elemental Spatial Density Neural Network Force Field (dubbed as Elemental-SDNNFF) with abundant atomic level environments as training data. Benefiting from the innovative architecture of the algorithm, sub-trillion atomic data can be integrated to train a single deep neural network for predicting complete phonon properties of >100,000 inorganic crystals spanning 63 elements in the periodic table. The workflow is shown in Fig.~\ref{fig:SDNNFF}. Such methodology enables us to discover promising thermal materials for various energy applications, including but not limited to thermoelectrics, interfacial thermal management, topological phonons for quantum information technology.

\subsection{Neurodevelopmental MRI Database}

Scientists using neuroimaging tools, such as MRI, NIRS, EEG, and MEG, use average
MRI templates for combining brain scans across different participants, locating brain structures
, segmenting priors for identification of cortical tissues, and identifying the relation between scalp locations and
the brain~\cite{sanchez2012neurodevelopmental}. However, use of these templates with pediatric, adolescent, and elderly
populations result in irregular registrations, misspecification of segmented data, and poor
fits for stereotaxic atlases. The “Neurodevelopmental MRI Database” was created to address
this issue. It provides MRI templates from 2 weeks through 89 years of age, and supporting
materials for neuroimaging. The initial formulation of the database only had access to 1.5T MRIs for the adult ages from 35 through 89 years. The current project will add 3.0T MRIs from four recent open-access MRI sites to
the adult ages. Led by Prof. John Richards, the project will refine the MRIs for wider use with
MRI processing programs and add these new materials to the NITRC open access site, and further enhance the current data and facilitate the sharing of this important resource. Our project has three aims.

\begin{figure*}[h]
  \centering
  \includegraphics[width=0.8\linewidth]{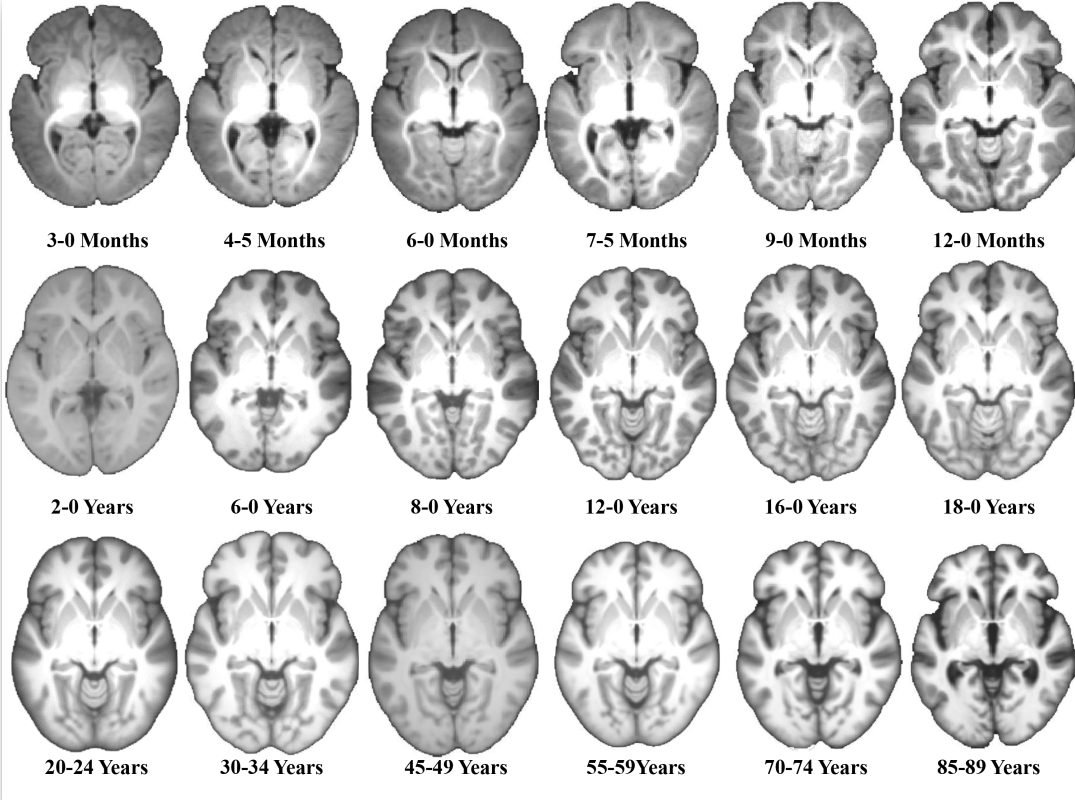}
  \caption{Brain templates for all ages.}
  \label{fig:brain}
\end{figure*}

Specific Aim 1—To construct average templates for adults through 89 years of age with
3.0T MRIs. The original database was constructed with participant MRIs from the University
of South Carolina McCausland Center for Brain Imaging, and the NIHPD, ABIDE, OASIS,
and IXI databases. Since that time, several new sharing sources have emerged. These first
were for infants and children, Pediatric Imaging, Neurocognition, and Genetic, Infant Brain Imaging Study, Baby CP and Developmental CP. These sources were used to update the database from 2 weeks through 20 years. The original adult average templates were from the OASIS and IXI databases. Five new sources
of scans with adult MRIs have become available since the initial database: Human Connectome Project, Human Connectome Project-aging, Cambridge Centre for Ageing and Neuroscience, Open Access Series of Imaging Studies, and the University of South Carolina Aging Brain Cohort study.
These sources comprise nearly 4,600 MRIs from participants ranging in age from 20 through
104 years. The average MRI templates and ancillary materials will be re-constructed from the existing data and these new scans. Sample templates are shown in Fig.~\ref{fig:brain}.

Specific Aim 2—To refine the averages to be applicable to a wider range of programs and
to enhance the ancillary materials. There are several refinements that were made to the
younger age data that will be applied to the adult data. One refinement will be to construct
the MRI volumes so they are easily used in a wider variety of neuroimaging programs. The
current set of atlases was based on the principles used by the FSL programming language.
This involved a specific volume orientation, radiological orientation, and
the structure of the volumes was enforced by the storage of the voxels in the image part of
the MRI volume. However, other neuroimaging processing systems allow more
flexibility in the image storage and rely more heavily on the header information for image
center, orientation, and affine translation. The current project will construct the average MRI
templates and associated materials to be more easily used in SPM12, Freesurfer and AFNI computer systems. A second refinement will be to develop ancillary materials that relate positions on the scalp to cortical areas inside
the head. Near Infrared Spectroscopy (“NIRS”) recordings use external scalp
sensors to record blood flow. There are photon migration models that simulate the flow of
the NIRS light inside the head. These models allow the specification of underlying cortical
areas for sensor placement, width between optode emitter/detector, and characteristics of
the head media. The individual data in the Neurodevelopmental MRI Database can be used
to develop probabilistic projection from scalp locations to defined stereotaxic cortical areas
for facilitating the use of the average MRI templates with fNIRS experiments.

Specific Aim 3—Store the data on the NITRC data archive. The NITRC site houses the ABIDE, IXI, and PING shared access datasets. We have over 350 sites and 1000 users who have access to the database from that site. The database materials have gained extensive public use and the templates will be in a “mature” state following this project. 

We heavily rely on Hyperion cluster to provide parallel computing to refine our average MRI templates and store template database.

\subsection{Moving Image Research Collections}
Moving Image Research Collections (MIRC) at the University Libraries has assembled a multidisciplinary team comprising faculty, students, and staff from the University’s Computer Vision Lab, film archive, Media Arts Program, and High-Performance Computing Group. The team's objective is to develop methodologies aimed at contextualizing and preserving the historical provenance of digitized surrogates of motion picture film from the United States Marine Corps Film Repository and other collections at MIRC. Our methodologies explore the potential of machine learning and neural networks for sophisticated analysis of image data generated from film stock. Additionally, we aim to establish a secure hub serving as a curated streaming video platform and a collaborative space for future research partners. 

\begin{figure*}[h]
  \centering
  \includegraphics[width=0.9\linewidth]{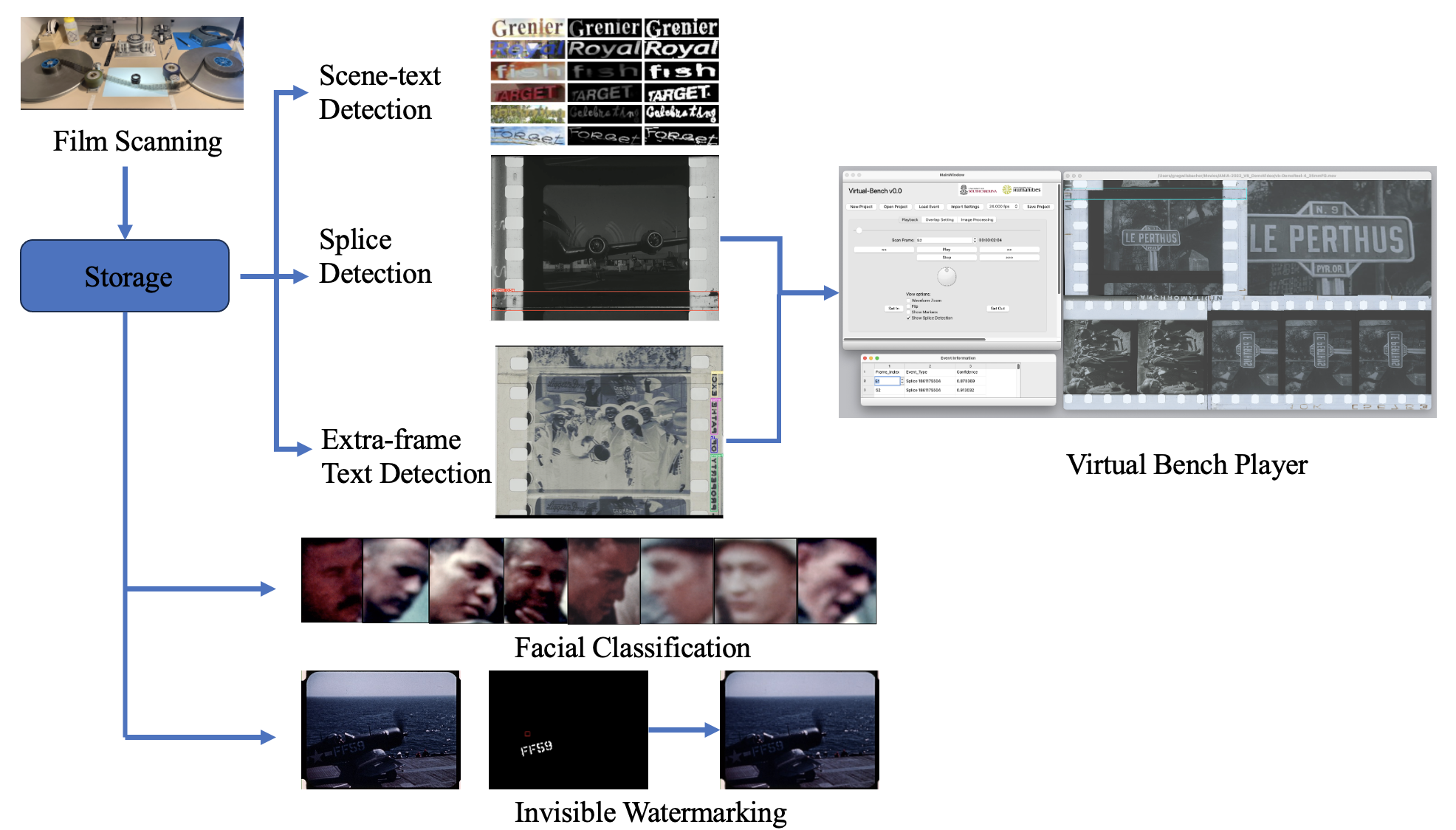}
  \caption{Pipeline of MIRC.}
  \label{fig:MIRC}
\end{figure*}

Under the leadership of Dr. Greg Wilsbacher, this project progresses through multiple stages. The pipeline is shown in Fig.~\ref{fig:MIRC}. Initially, professional technicians digitize the original film scans. To date, we have scanned 1,200 films, each up to 45 minutes long, with more to come. These videos are stored in AWS Glacier storage and the MIRC local SAN. Compressed videos are then transferred to the Hyperion cluster for processing. 

Subsequently, we address the rich contextual information present in these historical films, such as ship hull numbers, street signs, physical and printed-in splices, and extra-frame texts. We have developed an effective collection of deep-learning-based models capable of detecting text representations within complex contexts rapidly and accurately. Specifically, we trained two YOLO-based~\cite{jiang2022review} models to detect splices between frames and extra-frame texts, respectively. Additionally, we extended the textspotter~\cite{zhao2022background}~\cite{zhao2023commuspotter} algorithm to develop a new scene-text detection algorithm. This algorithm extracts hull numbers, building nameplates, and other challenging targets in historical videos. The outputs of these models are recorded as XML files and integrated as events into Virtual Bench the third stage of our project. Virtual Bench is an intelligent video player that allows users to concurrently display multiple frames and read/write/edit these events.

Furthermore, we explored video provenance methods using invisible digital watermarking based on deep-learning Encoder-Decoder algorithms. Additionally, human figures were extracted using facial extraction and classification methods to discover long-lost history.

Video processing can be extremely computationally intensive, particularly during deep-learning training. Hyperion enabled us to execute jobs in parallel across multiple nodes and multiple GPUs. Additionally, the Globus high-speed file transfer allowed us to upload and download videos within short timeframes.

\section{Conclusion}
In conclusion, the Hyperion HPC cluster represents a notable achievement in the USC research infrastructure, offering computational capabilities across diverse disciplines. Through customized construction and configuration, Hyperion provides researchers with the tools necessary to tackle complex computational challenges. The deployment of two machine learning models for predicting turnaround time helps researchers to optimize research workflows. Furthermore, the successful execution of three distinct research cases demonstrates the versatility and impact of Hyperion across various domains. As we continue to leverage Hyperion's power and flexibility, we are poised to push the boundaries of scientific inquiry and drive groundbreaking discoveries in the years to come.

\section{Acknowledgments}

We would like to express our sincere gratitude to all individuals and departments whose contributions were pivotal to the construction and operation of Hyperion. We extend our deepest appreciation to Provost Joan Gabel for her invaluable support, which played a crucial role in the success of this project. We are also thankful to the NSF MRI grant (award no. 2320292) for its generous support.

We wish to acknowledge the contributions of our colleagues and collaborators who provided invaluable insights and feedback throughout the duration of this project.

Lastly, we are grateful to the Division of Information Technology and the data center at USC for their unwavering administrative and facility support to the Hyperion cluster.

\bibliographystyle{ACM-Reference-Format}
\bibliography{hyperion}

@String{Computing = "Computing" }

@String{Computer = "{IEEE} Computer" }

@String{Springer = "Springer-Verlag" }

@inproceedings{tanash2021ampro,
  title={AMPRO-HPCC: A Machine-Learning Tool for Predicting Resources on Slurm HPC Clusters},
  author={Tanash, Mohammed and Andresen, Daniel and Hsu, William},
  booktitle={ADVCOMP... the... International Conference on Advanced Engineering Computing and Applications in Sciences},
  volume={2021},
  pages={20},
  year={2021},
  organization={NIH Public Access}
}

@inproceedings{bottum2013condo,
  title={The condo of condos},
  author={Bottum, James B and Marinshaw, Ruth and Neeman, Henry and Pepin, James and von Oehsen, J Barr},
  booktitle={Proceedings of the Conference on Extreme Science and Engineering Discovery Environment: Gateway to Discovery},
  pages={1--2},
  year={2013}
}

@article{palmer2015open,
  title={Open XDMoD: A tool for the comprehensive management of high-performance computing resources},
  author={Palmer, Jeffrey T and Gallo, Steven M and Furlani, Thomas R and Jones, Matthew D and DeLeon, Robert L and White, Joseph P and Simakov, Nikolay and Patra, Abani K and Sperhac, Jeanette and Yearke, Thomas and others},
  journal={Computing in Science \& Engineering},
  volume={17},
  number={4},
  pages={52--62},
  year={2015},
  publisher={IEEE}
}

@article{sanchez2012neurodevelopmental,
  title={Neurodevelopmental MRI brain templates for children from 2 weeks to 4 years of age},
  author={Sanchez, Carmen E and Richards, John E and Almli, C Robert},
  journal={Developmental psychobiology},
  volume={54},
  number={1},
  pages={77--91},
  year={2012},
  publisher={Wiley Online Library}
}

@article{rodriguez2023unlocking,
  title={Unlocking phonon properties of a large and diverse set of cubic crystals by indirect bottom-up machine learning approach},
  author={Rodriguez, Alejandro and Lin, Changpeng and Shen, Chen and Yuan, Kunpeng and Al-Fahdi, Mohammed and Zhang, Xiaoliang and Zhang, Hongbin and Hu, Ming},
  journal={Communications Materials},
  volume={4},
  number={1},
  pages={61},
  year={2023},
  publisher={Nature Publishing Group UK London}
}

@article{rodriguez2023million,
  title={Million-scale data integrated deep neural network for phonon properties of heuslers spanning the periodic table},
  author={Rodriguez, Alejandro and Lin, Changpeng and Yang, Hongao and Al-Fahdi, Mohammed and Shen, Chen and Choudhary, Kamal and Zhao, Yong and Hu, Jianjun and Cao, Bingyang and Zhang, Hongbin and others},
  journal={npj Computational Materials},
  volume={9},
  number={1},
  pages={20},
  year={2023},
  publisher={Nature Publishing Group UK London}
}

@article{zhao2023commuspotter,
  title={CommuSpotter: Scene Text Spotting with Multi-Task Communication},
  author={Zhao, Liang and Wilsbacher, Greg and Wang, Song},
  journal={Applied Sciences},
  volume={13},
  number={23},
  pages={12540},
  year={2023},
  publisher={MDPI}
}

@article{jiang2022review,
  title={A Review of Yolo algorithm developments},
  author={Jiang, Peiyuan and Ergu, Daji and Liu, Fangyao and Cai, Ying and Ma, Bo},
  journal={Procedia Computer Science},
  volume={199},
  pages={1066--1073},
  year={2022},
  publisher={Elsevier}
}

@inproceedings{zhao2022background,
  title={Background-Insensitive Scene Text Recognition with Text Semantic Segmentation},
  author={Zhao, Liang and Wu, Zhenyao and Wu, Xinyi and Wilsbacher, Greg and Wang, Song},
  booktitle={European Conference on Computer Vision},
  pages={163--182},
  year={2022},
  organization={Springer}
}

@article{rigatti2017random,
  title={Random forest},
  author={Rigatti, Steven J},
  journal={Journal of Insurance Medicine},
  volume={47},
  number={1},
  pages={31--39},
  year={2017},
  publisher={American Academy of Insurance Medicine 1700 Magnavox Way, Fort Wayne, IN 46804}
}


\end{document}